\documentclass{iaus}
\usepackage{graphics}

  \checkfont{eurm10}
  \iffontfound
    \IfFileExists{upmath.sty}
      {\typeout{^^JFound AMS Euler Roman fonts on the system,
                   using the 'upmath' package.^^J}%
       \usepackage{upmath}}
      {\typeout{^^JFound AMS Euler Roman fonts on the system, but you
                   dont seem to have the}%
       \typeout{'upmath' package installed. iaus.cls can take advantage
                 of these fonts,^^Jif you use 'upmath' package.^^J}%
      }
  \else
  \fi

  \checkfont{msam10}
  \iffontfound
    \IfFileExists{amssymb.sty}
      {\typeout{^^JFound AMS Symbol fonts on the system, using the
                'amssymb' package.^^J}%
       \usepackage{amssymb}%

      }{}
  \fi

  \IfFileExists{amsbsy.sty}
    {\typeout{^^JFound the 'amsbsy' package on the system, using it.^^J}%
     \usepackage{amsbsy}}
    {\providecommand\boldsymbol[1]{\mbox{\boldmath $##1$}}}

\newsavebox{\astrutbox}
\sbox{\astrutbox}{\rule[-5pt]{0pt}{20pt}}

\title[The Interplay among Black Holes, Stars and ISM in Galactic 
       Nuclei]{Stellar population properties in the nuclei and bulges
            of nearby lenticular galaxies}

\author[O. K. Sil'chenko]%
{Olga K. Sil'chenko$^1$}

\affiliation{$^1$Sternberg Astronomical Institute, Moscow State University,
University av. 13, Moscow 119992, Russia, email: olga@sai.msu.su}

\pubyear{2004}
\volume{222}
\pagerange{1--8}
\date{?? and in revised form ??}
\jname{The Interplay among Black Holes, Stars and ISM \\in Galactic Nuclei}
\editors{Th. Storchi Bergmann, L.C. Ho \& H.R. Schmitt, eds.}
\begin{document}

\maketitle

\begin{abstract}
The properties of stellar populations in the centers of nearby
lenticular galaxies are investigated by means of 2D spectroscopy.
All the galaxies are divided into 4 groups depending on the environment
type; every subsample contains more than 10 galaxies. Clear
distinctions between the mean stellar ages and abundance ratios both for
the nuclei and for the bulges of the S0s in the different
environments are found.
\end{abstract}

 During 1996--2003 we have obtained 2D spectroscopic data
in the spectral range of 4600--5400 \AA\ for 54 nearby lenticular galaxies
over various types of environments, namely, for the field galaxies, for the
brightest group galaxies (`group centers') and secondary group members, and
for the cluster lenticular galaxies. The observations have been made with
the Multi-Pupil Spectrograph of the 6m telescope (MPFS). Here we compare
the stellar population properties in the nuclei and in the `bulges' of these
galaxies, the term `bulges' being ascribed to the rings between
$R=3^{\prime \prime}$ and $R=7^{\prime \prime}$. The stellar population
abundances and ages are determined by comparing measured Lick indices
Mgb, Fe5270, Fe5335, and H$\beta$ to the evolutionary synthesis
models of \cite{thomod}. As we have reported earlier
(\cite[Sil'chenko 1993]{me93}, \cite[Sil'chenko 2002]{silmex}), a
significant part of nearby lenticular galaxies possess young stellar nuclei;
according to the present estimates, among field S0s and
group second- or third-ranked members 50\%\ galaxies and among group centers
and cluster members 25\%\ galaxies reveal mean age of the nuclear 
stellar population of less than 5~Gyr. A bulge is almost
always older than the nucleus. An effect of environments is seen both
for the nuclei and for the bulges: if we consider the clusters (Virgo and
Ursa Major) and group
centers as dense environments and the field and group periphery as sparse
environments, then in the dense environments the stellar
populations of S0s are in average older by 4--5 Gyr than in the sparse ones.

We have analysed carefully the index-index diagrams,
$\langle \mbox{Fe} \rangle \equiv$(Fe5270+Fe5335)/2 {\bf vs} Mgb and
H$\beta$ {\bf vs}
[MgFe]$\equiv \sqrt{\mbox{Mgb} \langle \mbox{Fe} \rangle}$,
for all four types of environments; the stellar absorption-line
index H$\beta$ has been corrected for the emission contamination
mostly by using the H$\alpha$ emission estimates from \cite{hofil97}.
We have come to the following particular conclusions:\\

\begin{enumerate}

\item{
In the dense environments, [Mg/Fe]$_{nuc}$ varies from $+0.2$ to $+0.4$
under $\mbox{[m/H]}\approx \mbox{const} \approx +0.3$. This result
implies very brief and very effective last star formation epochs
in the nuclei of the lenticulars in the dense environments.
}

\item{
In the sparse environments,
[Mg/Fe]$_{nuc} \approx \mbox{const} \approx +0.1 - +0.2$,
while $\mbox{[m/H]}$ varies from $-0.3$ to $+0.3$. The metallicity
variations seem to reflect a well-known mass-metallicity relation
for early-type galaxies.
}

\item{
The bulges within each type of environments represent a metallicity sequence
under [Mg/Fe]$\approx$const, exactly as the nuclei of the lenticulars
in the sparse environments, but the mean [Mg/Fe] falls when passing
from the dense environments to the sparse ones.
}

\item{
Among the nuclei of S0s of all types of environments one can meet mean stellar 
ages from 1~Gyr to 15~Gyr.
}

\item{
The bulges have mean stellar ages between 5 and 15~Gyr, the bulges
in the field lenticulars being significantly younger than in S0s of
other types of environments. However, we must note that in some galaxies
what we regard as the `bulges' may be in fact a sum of the bulges and of 
extended circumnuclear disks produced by secondary star formation bursts.
If a circumnuclear disk is close to face-one orientation, it cannot
be detected either by photometry nor by rotation.
}

\item{
Cumulative age distributions (Fig.~1) reveal a systematic shift between 
the stellar age distributions of both the nuclei and the bulges in the dense
and the sparse environments: the median ages of the nuclei are 8~Gyr in
the dense environments and 5.5~Gyr in the sparse ones, and the median ages
of the bulges are 14~Gyr and 9~Gyr, respectively.}

\end{enumerate}

These observational results must be compared with the predictions
of the cosmology-based scenario of galaxy formation which imply
environment dependence of the formation epoch similar to that
found by us in this work. However I feel that the absolute mean age
values for the nuclei and bulges of the nearby luminous S0s galaxies may be
in come contradiction with the prescriptions of the $\Lambda$CDM model. 
I would stress also the difference between the nuclei and the bulges
which appear to be quite different stellar subsystems, with absolutely
decoupled evolutions. The secondary star formation bursts are obviously
more frequent in the nuclei than over extended areas in the bulges.

\begin{figure}
\resizebox{\hsize}{!}{\includegraphics{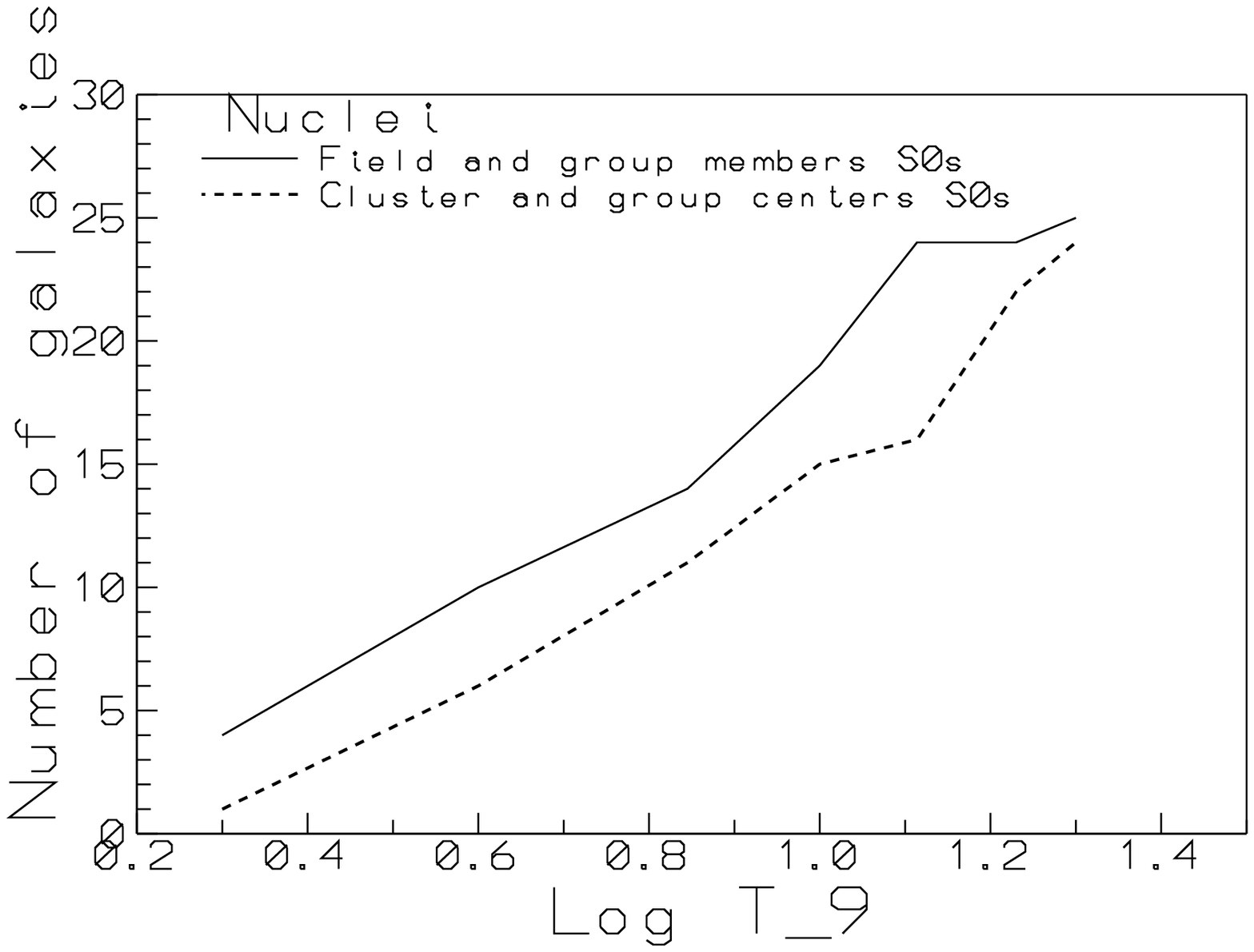}
\includegraphics{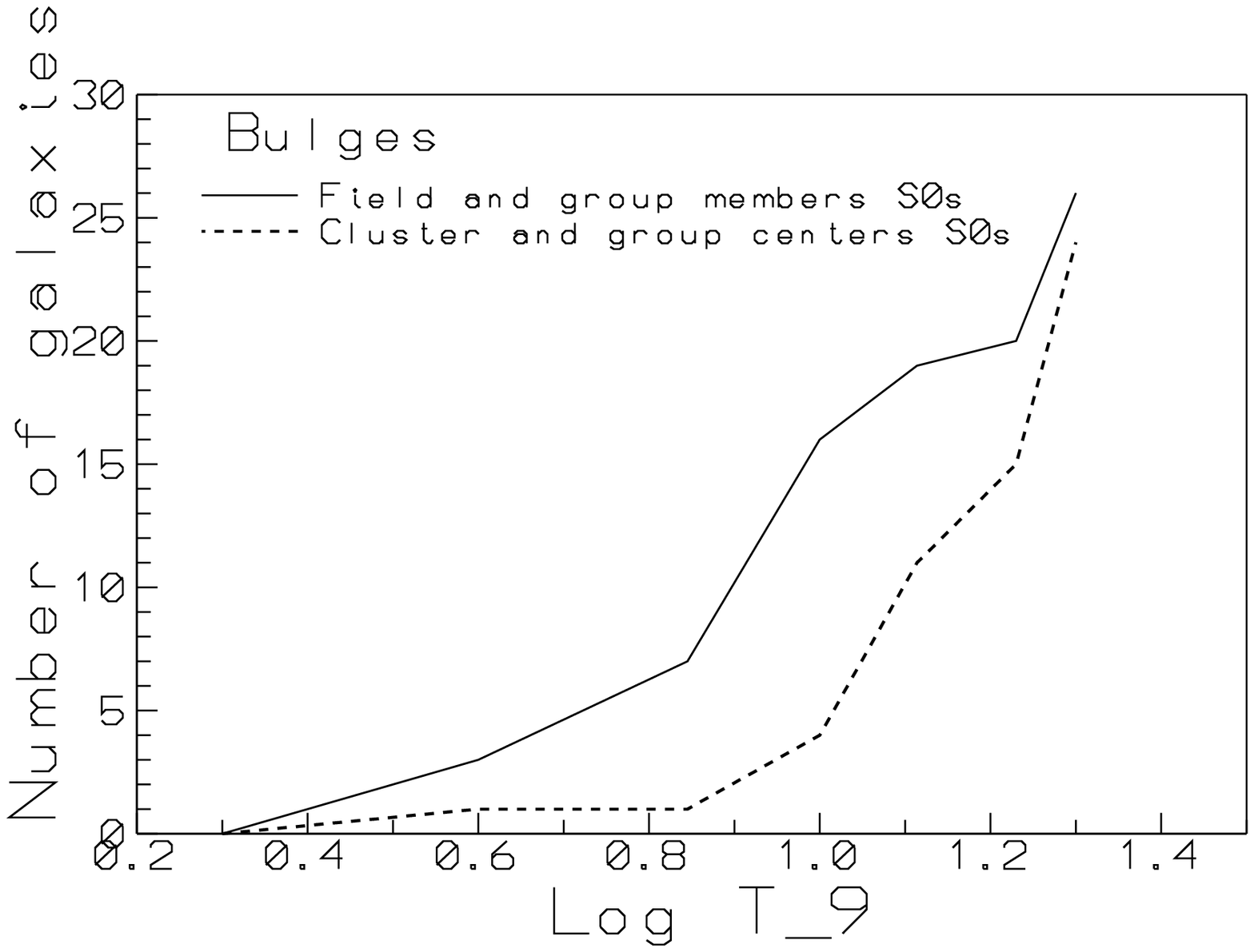}}
  \caption{Cumulative age distributions: a number of objects
  younger than abcissa which is $\lg T$ in Gyr.
    (\textit{a}) The stellar nuclei of the galaxies
    (\textit{b}) The bulges taken in the rings between
    $R=3^{\prime \prime}$ and $R=7^{\prime \prime}$.}
\end{figure}

\begin{acknowledgments}
I thank my collaborators V. L. Afanasiev, A. N. Burenkov, A. V. Moiseev,
and V. V. Vlasyuk. The study of young nuclei in lenticular galaxies
has been supported by the Russian Foundation for Basic Research 
(grant 01-02-16167)
and by the Federal Scientific-Technical Program -- contract
of the Science Ministry of Russia no.40.022.1.1.1101.

\end{acknowledgments}

\end{document}